\documentclass[twocolumn,showpacs,pra,floatfix]{revtex4}

\usepackage{graphicx}
\usepackage{bm}
\usepackage{psfrag}
\usepackage{amsmath}
\usepackage{amssymb}
\usepackage{latexsym}
\usepackage{exscale}

\newcommand{\vect}[1]{\bm{#1}}
\newcommand{\ten}[1]{\mbox{\textbf{
{\textsf{#1}}}}}

\newcommand{\veczero}{\mbox{\textbf{\textit{0}}}}
\newcommand{\tenszero}{\mbox{\textbf{\textsf{0}}}}
\newcommand{\sprod}{\!\cdot\!}
\newcommand{\tprod}{}
\newcommand{\vprod}{\!\times\!}
\newcommand{\trace}{\operatorname{tr}}
\newcommand{\trans}{\mathsf{T}}
\newcommand{\dif}{\mathrm{d}}
\newcommand{\mi}{\mathrm{i}}
\newcommand{\me}{\mathrm{e}}

\begin{document}

\title{Body-assisted dispersion potentials of diamagnetic atoms}

\date{\today}

\author{Stefan Yoshi Buhmann}
\affiliation{Quantum Optics and Laser Science, Blackett
Laboratory, Imperial College London, Prince Consort Road, London SW7
2AZ, United Kingdom}

\author{Hassan Safari}
\affiliation{Kerman, Iran}

\author{Stefan Scheel}
\affiliation{Institut f\"ur Physik, Universit\"at Rostock,
Universit\"atsplatz 3, D-18051 Rostock, Germany}
\affiliation{Quantum Optics and Laser Science, Blackett Laboratory,
Imperial College London, Prince Consort Road, London SW7 2AZ, United
Kingdom}

\author{A.~Salam}
\affiliation{Department of Chemistry, Wake Forest University,
Winston-Salem, NC 27109, USA}

\date{\today}

\begin{abstract}
We derive Casimir--Polder and van der Waals potentials of one or
two atoms with diamagnetic properties in an arbitrary environment of
magnetoelectric bodies. The calculations are based on macroscopic
quantum electrodynamics and leading-order perturbation theory. For
the examples of an atom and a perfect mirror and two atoms in free
space we show that diamagnetic dispersion potentials have the same
sign as their electric counterparts, but can exhibit quite different
distance dependences.
\end{abstract}

\maketitle

%
\section{Introduction}
\label{Sec1}

The issue of electrodynamics in magnetodielectrics is a long-standing
one \cite{Landau60}. These range from the form and propagation of
Maxwell fields in such media, to their absorption and emission
characteristics, as well as to the numerous other optical properties
that they exhibit. One curious aspect, recently confirmed by
fabrication of periodic arrays of thin metallic wires
\cite{Smith00,Shelby01,Marques02}, is the existence of left-handed
materials, so-called because the electric, magnetic and direction of
propagation vectors form a left-handed triad, resulting in a
metamaterial with negative refractive index. Another interesting
aspect is the nature of inter-particle interactions in media with
electrical permittivity, magnetic permeability, or both. This has its
origins in the work of Casimir \cite{Casimir48}, subsequently spawning
a wide variety of related phenomena categorized under Casimir-like
effects \cite{Milton04,Bordag09,Dalvit11,Hertlein08}, and including
the Casimir--Polder (CP) dispersion potential \cite{CasimirPolder48}.

The understanding and computation of van der Waals forces (vdW) for
chemical, physical and biological systems is of fundamental importance
and widespread interest as they manifest in vacuum, gaseous or
condensed phases. These vary from semi-classical treatments
yielding the London dispersion formula, to quantum field theoretic
approaches where radiative effects are properly accounted for and
result in Casimir energy shifts.  Perhaps the best-known example of a
fundamental theory that automatically includes retardation is quantum
electrodynamics (QED). Two often-used versions include macroscopic
\cite{0696,0853,Philbin10,Horsley11} and microscopic
\cite{Craig84,Salam10} QED, their prefixes aptly describing their
general range of applicability. In the former variant, the
body-assisted electromagnetic field is evaluated by quantizing the
radiation field and the dispersive and absorptive medium. This enables
the body-induced atomic energy shift to be computed, which is
interpreted as the potential of the force. From this, pair and
many-body interaction energies can also be calculated. This has been
successfully carried out for both electrically polarizable
\cite{Safari06} and paramagnetically susceptible systems \cite{0831}.
Effects of an intermediate medium can also be accounted for
\cite{0831,0739}.

Microscopic Coulomb gauge QED constructed via quantization of the free
electromagnetic field has also been used successfully to evaluate
atom--field and atom--atom interactions in a vacuum. In the latter
situation this has included re-calculation of the CP potential, and
its extension to higher multipole moment contributions, such as
magnetic dipole, electric quadrupole and diamagnetic couplings
\cite{Mavroyannis62,Thirunamachandran88,Jenkins94,Salam96,Salam00,
Salam00b,Babb10}. Unexpected results ensue such as the discriminatory
nature of the interaction between two chiral (optically active)
molecules; the repulsive form of the ground state dispersion energy
shift between an electrically polarizable atom and a paramagnetically
susceptible one; and that in the near-zone, the electric--diamagnetic
contribution to this potential is larger than the corresponding limit
arising from the electric--paramagnetic term \cite{Jenkins94}.

In this paper a systematic study is performed on the CP potentials of
a diamagnetic atom and its vdW potentials with another, and with an
atom that is either electrically or paramagnetically polarizable in a
medium comprised of magnetodielectric bodies using macroscopic QED
theory. Interest in contributions arising from diamagnetic coupling
has been due to their importance when computing highly accurate
potentials for alkali metal atom dimers \cite{Marinescu99}, where it
has been found that the electric--diamagnetic and
paramagnetic--diamagnetic terms can be larger than the
electric--paramagnetic and paramagnetic--paramagnetic energy shifts.
The article is organized as follows. Section~\ref{Sec2} briefly
describes the quantized body-assisted Maxwell field operators in the
medium, their expression in terms of the Green's-function solution of
the Helmholtz equation, and the writing of the total system
Hamiltonian. Electric and total magnetic contributions to the single
atom CP shift, and for an atom placed in front of a perfectly
reflecting mirror are obtained in Sec.~\ref{Sec3}. Presented in
Sec.~\ref{Sec4} are diagrammatic perturbation theory results for the
dispersion pair potential when at least one of the atoms is
diamagnetic. Corresponding free-field interaction energies are also
given, and these are briefly compared with results obtained via
microscopic QED. In all of the cases examined, the asymptotic limits
of the energy shift in the near- and far-zones corresponding to
non-retarded and retarded regimes are found, and compared with
previously obtained limiting forms for electric, paramagnetic, and
electric--paramagnetic interactions. A short summary is given in
Sec.~\ref{Sec5}.
%
%
\section{Quantisation scheme}
\label{Sec2}
Consider one or two atoms (or molecules) with internal Hamiltonians
($\xi=A,B$)
\begin{equation}
\label{eq1}
\hat{H}_\xi
 =\sum_n E_\xi^n|n_\xi\rangle\langle n_\xi|
\end{equation}
which are placed at positions $\vect{r}_\xi$ within an arbitrary
arrangement of linearly, locally and isotropically responding
magnetoelectric bodies, described by a Kramers-Kronig consistent
permittivity $\varepsilon(\vect{r},\omega)$ and permeability
$\mu(\vect{r},\omega)$. Upon introducing bosonic variables
$\hat{\vect{f}}_\lambda^\dagger(\vect{r},\omega)$ and
$\hat{\vect{f}}_\lambda(\vect{r},\omega)$, which are creation and
annihilation operators for the elementary electric ($\lambda=e$) and
magnetic ($\lambda=m$) excitations of the system of bodies and
electromagnetic field and obey the bosonic commutation relations
\begin{equation}
\label{comm1}
[\hat{\vect{f}}_\lambda(\vect{r},\omega),
\hat{\vect{f}}_{\lambda'}^\dagger(\vect{r}',\omega')] =\bm{\delta} 
(\vect r-\vect r')\delta_{\lambda \lambda'}\delta(\omega-\omega')
\end{equation}
and
\begin{equation}
\label{comm2}
[\hat{\vect{f}}_\lambda(\vect{r},\omega),\hat{\vect{f}}_{\lambda'}
(\vect{r}',\omega')] =[\hat{\vect{f}}_\lambda^\dagger(\vect{r},
\omega),\hat{\vect{f}}_{\lambda'}^\dagger(\vect{r}',\omega')] 
=\ten{0},
\end{equation}
the body--field Hamiltonian takes the form \cite{0853}
\begin{equation}
\label{eq2}
\hat{H}_F
 =\sum_{\lambda=e,m}
 \int\dif^3r\int_0^\infty\dif\omega\,\hbar\omega
 \hat{\vect{f}}_\lambda^\dagger(\vect{r},\omega)
 \sprod\hat{\vect{f}}_\lambda(\vect{r},\omega).
\end{equation}
The ground-state of $\hat{H}_F$ can obviously be defined by
\begin{equation}
\label{eq2b}
\hat{\vect{f}}_\lambda(\vect{r},\omega)|\{0\}\rangle
=\veczero\quad\forall \lambda,\vect{r},\omega.
\end{equation}
Within the multipolar coupling scheme, the interaction of each atom
with the body-assisted electromagnetic field is given by \cite{0008}
\begin{equation}
\label{eq3}
\hat{H}_{\xi F}=\hat{H}_{\xi F}^e+\hat{H}_{\xi F}^p+\hat{H}_{\xi F}^d.
\end{equation}
Here the three terms are the electric, paramagnetic and diamagnetic
interactions,
\begin{eqnarray}
\label{elec-int}
\hat{H}_{\xi F}^e &=& -\hat{\vect{\mu}}_\xi\sprod
\hat{\vect{E}}(\vect{r}_\xi),\\
\label{mag-int}
\hat{H}_{\xi F}^p &=&
-\hat{\vect{m}}_\xi\sprod\hat{\vect{B}}(\vect{r}_\xi),
\\
\label{dia-int} 
\hat{H}_{\xi F}^d &=&\sum\limits_{\alpha\in \xi}
\frac{q_\alpha^2}{8m_\alpha}\bigl[\hat{\bar{\vect{r}}}_\alpha\vprod
\hat{\vect{B}}(\vect{r}_\xi)\bigr]^2,
\end{eqnarray}
with $\hat{\vect{\mu}}_\xi$ and $\hat{\vect{m}}_\xi$ being the
respective atomic electric and magnetic dipole operators and
$q_\alpha$, $m_\alpha$ and $\hat{\bar{\vect{r}}}_\alpha$ denoting the
charges, masses and positions relative to the center of mass of the
particles contained in the atoms. Note that electric quadrupole and
even octupole contributions can easily be included in the formalism;
they have recently been shown to affect the CP potential of Rydberg
atoms close to surfaces \cite{0892}.

Introducing the atomic diamagnetisability operator as
\begin{equation}
\hat{\bm{\beta}}_\xi^d =-\sum_{\alpha \in \xi} \frac{q_\alpha^2}
{4m_\alpha}\, (\hat{\bar{\vect{r}}}_\alpha^2\ten{I} 
-\hat{\bar{\vect{r}}}_\alpha\hat{\bar{\vect{r}}}_\alpha ),
\end{equation}
and using the identity
$[\vect{a}\vprod\vect{b}]^2 = \vect{b}\cdot(\vect{a}^2\ten{I} - 
\vect{a}\vect{a})\cdot\vect{b}$, the diamagnetic interaction
Hamiltonian may be cast in the form
\begin{equation}
\label{mag}
\hat{H}_{\xi F}^d = -\tfrac{1}{2}\hat{\vect{B}}
(\vect{r}_\xi)\sprod\hat{\bm{\beta}}_\xi^d \sprod
\hat{\vect{B}}(\vect{r}_\xi).
\end{equation}
The ground-state diamagnetisability of an atom is given by the
expectation value
\begin{eqnarray}
\label{eq21}
\bm{\beta}_\xi^d&\equiv&
\langle\hat{\bm{\beta}}_\xi^d\rangle
=-\sum_{\alpha\in\xi}\frac{q_\alpha^2}{4m_\alpha}\,
 \langle 0_\xi| \hat{\bar{\vect{r}}}_\alpha^2\ten{I}
 -\hat{\bar{\vect{r}}}_\alpha\tprod\hat{\bar{\vect{r}}}_\alpha
 |0_\xi\rangle\nonumber\\
&=&-\sum_{\alpha\in\xi}\frac{q_\alpha^2}{6m_\alpha}\,
 \langle\hat{\bar{\vect{r}}}_\alpha^2\rangle\ten{I}
\equiv\beta_\xi^d\ten{I},
\end{eqnarray}
where the second line holds for isotropic atoms.

The total Hamiltonian of the atom(s) interacting with the
electromagnetic field in the presence of the bodies takes the form
\begin{equation}
\label{eq4}
\hat{H}=\sum_{\xi=A,B}\hat{H}_\xi+\hat{H}_F
 +\sum_{\xi=A,B}\hat{H}_{\xi F}.
\end{equation}
The electric and  magnetic field operators can be expanded in terms of
the bosonic operators
$\hat{\vect{f}}_\lambda^\dagger(\vect{r},\omega)$ and
$\hat{\vect{f}}_\lambda(\vect{r},\omega)$ as
\begin{eqnarray}
\label{eq5}
\hat{\vect{E}}(\vect{r})
&\!=&\!\sum_{\lambda={e},{m}}
 \int\dif^3r'\int_0^{\infty}\dif\omega\,
 \ten{G}_\lambda(\vect{r},\vect{r}',\omega)
 \sprod\hat{\vect{f}}_\lambda(\vect{r}',\omega)
 \nonumber\\
&\!&\!+\operatorname{H.c.}\,,\\
\label{eq7}
\hat{\vect{B}}(\vect{r})
&\!=&\!\sum_{\lambda={e},{m}}
 \int\dif^3r'\int_0^{\infty}\frac{\dif\omega}{\mi\omega}\,
 \vect{\nabla}\vprod
 \ten{G}_\lambda(\vect{r},\vect{r}',\omega)
 \sprod\hat{\vect{f}}_\lambda(\vect{r}',\omega)
 \nonumber\\
&\!&\!+\operatorname{H.c.}
\end{eqnarray}
The expansion coefficients $\ten{G}_\lambda$ are related
to the classical Green tensor $\ten{G}$ by
\begin{align}
\label{eq8}
&\ten{G}_e(\vect{r},\vect{r}',\omega)
 =\mi\,\frac{\omega^2}{c^2}
 \sqrt{\frac{\hbar}{\pi\varepsilon_0}\,
 \operatorname{Im}\varepsilon(\vect{r}',\omega)}\,
 \ten{G}(\vect{r},\vect{r}',\omega),\\
\label{eq9}
&\ten{G}_m(\vect{r},\vect{r}',\omega)
 =\mi\,\frac{\omega}{c}
 \sqrt{\frac{\hbar}{\pi\varepsilon_0}\,
 \frac{\operatorname{Im}\mu(\vect{r}',\omega)}
 {|\mu(\vect{r}',\omega)|^2}}
 \bigl[\vect{\nabla}'
 \!\!\times\!\ten{G}(\vect{r}',\vect{r},\omega)
 \bigr]^{\trans};
\end{align}
and the Green tensor is the unique solution to the Helmholtz
equation
\begin{equation}
\label{eq10}
\left[\bm{\nabla}\times
\frac{1}{\mu(\vect{r},\omega)}\bm{\nabla}\times
 \,-\,\frac{\omega^2}{c^2}\,\varepsilon(\vect{r},\omega)\right]
 \ten{G}(\vect{r},\vect{r}',\omega)
 =\bm{\delta}(\vect{r}-\vect{r}')
\end{equation}
together with the boundary condition
\begin{equation}
\label{eq11}
\ten{G}(\vect{r},\vect{r}',\omega)\to \tenszero
\quad\mbox{for }|\vect{r}-\vect{r}'|\to\infty.
\end{equation}
Just like the permittivity and permeability, the Green tensor is an
analytic function in the upper half of the complex frequency plane 
and fulfils the Schwarz reflection principle
\begin{equation}
\label{eq12}
\ten{G}(\vect{r},\vect{r}',-\omega^\ast) 
 =\ten{G}^\ast(\vect{r},\vect{r}',\omega).
\end{equation}
In addition, it obeys the Onsager--Lorentz reciprocity
\begin{equation}
\label{eq13}
\ten{G}(\vect{r}',\vect{r},\omega) 
 =\ten{G}^\trans(\vect{r},\vect{r}',\omega)
\end{equation}
and the useful integral relation
\begin{multline}
\label{eq14}
\sum_{\lambda={e},{m}}\int\dif^3 s\,
 \ten{G}_\lambda(\vect{r},\vect{s},\omega)\!\cdot\!
 \ten{G}^{\ast\trans}_\lambda(\vect{r}',\vect{s},\omega)\\
=\frac{\hbar\mu_0}{\pi}\,\omega^2\operatorname{Im}
 \ten{G}(\vect{r},\vect{r}',\omega)
\end{multline}
holds.
%
%
\section{Casimir--Polder potential of a single atom}
\label{Sec3}
In this section, we calculate the CP force on a single atom in
the presence of magnetoelectric bodies, discarding the label $A$
wherever possible. According to Casimir and Polder, the force can be
derived from the associated CP potential $U(\vect{r}_A)$
\begin{equation}
\label{e15}
\vect{F}=-\vect{\nabla}U(\vect{r}_A),
\end{equation}
which in turn can be identified as the position-dependent part of the
energy shift $\Delta E$ due to the atom--field coupling
\begin{equation}
\label{e16}
U(\vect{r}_A)=\Delta E(\vect{r}_A).
\end{equation}
%
%
\subsection{Perturbation theory}
\label{Sec3.1}
We assume the atom--field system to be prepared in the uncoupled
ground state $|0_A\rangle|\{0\}\rangle$ and calculate the energy shift
due to the atom--field coupling within leading-order perturbation
theory. Let us first study a purely diamagnetic atom, whose CP
potential is due to the first-order energy shift associated with the
diamagnetic part of the atom--field interaction~(\ref{mag}),
\begin{multline}
\label{eq17}
\Delta E=\langle\{0\}|\langle 0_A|
-\tfrac{1}{2}\,\hat{\vect{B}}(\vect{r}_A)\sprod\hat{\bm{\beta}}^d
\sprod
\hat{\vect{B}}(\vect{r}_A)|0_A\rangle|\{0\}\rangle.
\end{multline}
Normally, CP energy shifts only arise in second order perturbation
theory. However, due to the diamagnetic interaction Hamiltonian being
quadratic in the magnetic induction field, already the first
perturbation order contributes. It can be easily evaluated by using
the magnetic-field expression~(\ref{eq7}), the bosonic commutation
relations (\ref{comm1}) and (\ref{comm2}), and the integral
relation~(\ref{eq14}), resulting in
\begin{equation}
\label{eq18}
\Delta E=\frac{\hbar\mu_0}{2\pi}
 \int_0^\infty\dif\omega\trace\bigl[{\bm{\beta}}^d
  \sprod
 \operatorname{Im}\ten{G}_{mm}(\vect{r}_A,\vect{r}_A,\omega)\bigr]
 \end{equation}
with
\begin{equation}
\label{L}
\ten{G}_{mm} (\vect r,\vect r',\omega) = \vect{\nabla}\vprod\ten{G}
(\vect{r},\vect{r}',\omega)
 \vprod\overleftarrow{\vect{\nabla}}'.
\end{equation}

For an atom in a free-space region, the CP potential can be extracted
from this energy shift by separating the Green tensor into its bulk
and scattering parts,
\begin{equation}
\label{eq19}
\ten{G}(\vect{r},\vect{r}',\omega)
=\ten{G}^{(0)}(\vect{r},\vect{r}',\omega)
 +\ten{G}^{(1)}(\vect{r},\vect{r}',\omega),
\end{equation}
and discarding the constant energy shift associated with the
translationally invariant bulk part by making the replacement
$\ten{G}\mapsto\ten{G}^{(1)}$. The result can be simplified
by converting the integral over real frequencies to another along the 
imaginary axis in the complex frequency plane. To this end, we first 
write
\begin{equation}
\int_0^\infty \dif \omega \operatorname{Im} \ten{G}(\vect{r},
\vect{r}',\omega) = \operatorname{Im}\int_0^\infty \dif \omega  
\ten{G}(\vect{r},\vect{r}',\omega).
\end{equation}
The integral on the right hand side can be replaced by an integral
along the positive imaginary axis ($\omega$ $\!\mapsto$ $\!\mi \xi$) 
plus a vanishing integral along infinite quarter-circles via Cauchy's 
theorem. Using the fact that $\ten G(\mi \xi)$ is 
real-valued for a real $\xi$, as can be inferred from Schwarz
reflection principle (\ref{eq12}), one finds
\begin{align}
\label{eq22}
U_d(\vect{r}_A)
 =&\frac{\hbar\mu_0}{2\pi}
 \int_0^\infty\!\!\!\dif\xi\,\trace\bigl[\bm{\beta}^d
 \sprod\ten{G}_{mm}^{(1)}(\vect{r}_A,\vect{r}_A,\mi\xi)
\bigr]\nonumber\\
=&\frac{\hbar\mu_0}{2\pi}
 \int_0^\infty\!\!\!\dif\xi\,\beta^d\trace
 \ten{G}_{mm}^{(1)}(\vect{r}_A,\vect{r}_A,\mi\xi).
\end{align}
The second line in Eq.~(\ref{eq22}) holds for isotropic
atoms. For an atom with an additional nontrivial electric and
paramagnetic response, the respective electric and paramagnetic
interactions in the coupling Hamiltonian~(\ref{eq3}) also need to be
taken into account. As shown previously, they give rise to
second-order energy shifts such that the electric and paramagnetic CP
potentials are given by \cite{0831}
\begin{eqnarray}
\label{eq23}
U_e(\vect{r}_A)
 &=&
\frac{\hbar}{2\pi\varepsilon_0}
 \!\int_0^\infty\!\!\!\dif\xi\,
 \trace\bigl[\bm{\alpha}(\mi\xi)\sprod
 \ten{G}_{ee}^{(1)}(\vect{r}_A,\vect{r}_A,\mi\xi)\bigr]\nonumber\\
&=&\frac{\hbar}{2\pi\varepsilon_0}\!\int_0^\infty\!\!\!\dif\xi\,
 \alpha(\mi\xi)
 \trace\ten{G}_{ee}^{(1)}(\vect{r}_A,\vect{r}_A,\mi\xi),\\
\label{eq24}
U_p(\vect{r}_A)
 &=&\frac{\hbar\mu_0}{2\pi}
 \!\int_0^\infty\!\!\!\dif\xi\,\trace\bigl[\bm{\beta}^p(\mi\xi)
 \sprod \ten{G}_{mm}^{(1)}(\vect{r}_A,\vect{r}_A,\mi\xi)\bigr]
 \nonumber\\
&=&\frac{\hbar\mu_0}{2\pi}
 \!\int_0^\infty\!\!\!\dif\xi\,\beta^p(\mi\xi)\trace
  \ten{G}_{mm}^{(1)}(\vect{r}_A,\vect{r}_A,\mi\xi)\quad
\end{eqnarray}
with
\begin{equation}
\ten{G}_{ee}(\vect{r},\vect{r},\omega)= -\frac{\omega^2}{c^2}
\ten{G}(\vect{r},\vect{r},\omega)
\end{equation}
and
\begin{eqnarray}
\label{eq25}
\bm{\alpha}(\omega)
&=&\lim_{\epsilon\to 0}\frac{2}{\hbar}\sum_k
 \frac{\omega_A^{k}
\vect{\mu}_A^{0k}\tprod\vect{\mu}_A^{k0}}
 {(\omega_A^k)^2\!-\!\omega^2\!-\!\mi\omega\epsilon}\nonumber\\
&=&\lim_{\epsilon\to 0}\frac{2}{3\hbar}\sum_k
 \frac{\omega_A^k
|\vect{\mu}_A^{0k}|^2}
 {(\omega_A^k)^2\!-\!\omega^2\!-\!\mi\omega\epsilon}\,\ten{I}
 =\alpha(\omega)\ten{I},\\
\label{eq26}
\bm{\beta}^p(\omega)
&=&\lim_{\epsilon\to 0}\frac{2}{\hbar}\sum_k
 \frac{\omega_A^k
\vect{m}_A^{0k}\tprod\vect{m}_A^{k0}}
 {{\omega_A^k}^{\!2}\!-\!\omega^2\!-\!\mi\omega\epsilon}\nonumber\\
&=&\lim_{\epsilon\to 0}\frac{2}{3\hbar}\sum_k
 \frac{\omega_A^k|\vect{m}_A^{0k}|^2}
 {(\omega_A^k)^2\!-\!\omega^2\!-\!\mi\omega\epsilon}\,\ten{I}
 =\beta^p(\omega)\ten{I}\qquad
\end{eqnarray}
[$\omega_A^k=(E_A^k-E_A^0)/\hbar$, 
$\vect{\mu}_A^{0k}=\langle 0|\hat{\vect{\mu}}_A|k\rangle$,
$\vect{m}_A^{0k}=\langle 0|\hat{\vect{m}}_A|k\rangle$] denoting the
polarisability and paramagnetisability of the atom, respectively.
Introducing the total magnetisability
\begin{eqnarray}
\label{eq27}
\bm{\beta}(\omega)
 &=&\bm{\beta}^p(\omega)+\bm{\beta}^d\nonumber\\
 &=&[\beta^p(\omega)+\beta^d]\ten{I}
 =\beta(\omega)\ten{I},
\end{eqnarray}
the magnetic part of the CP potential reads
\begin{eqnarray}
\label{eq28}
U_m(\vect{r}_A)&=&U_p(\vect{r}_A)+U_d(\vect{r}_A)\nonumber\\
 &=&\frac{\hbar\mu_0}{2\pi}
 \!\int_0^\infty\!\!\!\dif\xi\,\trace\bigl[\bm{\beta}(\mi\xi)
 \sprod
\ten{G}_{mm}^{(1)}
(\vect{r}_A,\vect{r}_A,\mi\xi)
 \bigr]\nonumber\\
&=&\frac{\hbar\mu_0}{2\pi}
 \!\int_0^\infty\!\!\!\dif\xi\,\beta(\mi\xi)\trace\bigl[
\ten{G}_{mm}^{(1)}
(\vect{r}_A,\vect{r}_A,\mi\xi)
\bigr]
\end{eqnarray}
and the total CP potential is given by
\begin{equation}
\label{eq29}
U(\vect{r}_A)=U_e(\vect{r}_A)+U_m(\vect{r}_A).
\end{equation}

We have thus generalised previous results for the CP potential of an
atom with electric and paramagnetic properties to one that also
exhibits nontrivial diamagnetic properties. It is found that despite
the different interaction terms and perturbative orders (first order
instead of second order), the extension to a diamagnetic atom can be
obtained formally by including the diamagnetic contribution in the
magnetisability, $\bm{\beta}^p(\omega)\mapsto\bm{\beta}(\omega)%
=\bm{\beta}^p(\omega)+\bm{\beta}^d$. In particular, the local-field
corrected potentials for atoms embedded in a medium as derived in
Ref.~\cite{0831} remain valid with this replacement.

By introducing $\bm{\alpha}^e(\omega)$ $\!=$ $\!\bm{\alpha}(\omega)$,
$\bm{\alpha}^m(\omega)$ $\!=$ $\!\bm{\beta}(\omega)/c^2$, the
electric and magnetic parts of the CP potential can be given in the
compact notation
\begin{align}
\label{ul}
U_\lambda(\vect{r}_A)
 =&\frac{\hbar}{2\pi\varepsilon_0}
 \!\int_0^\infty\!\!\!\dif\xi
 \trace\bigl[\bm{\alpha}^\lambda(\mi\xi)\sprod
 \ten{G}_{\lambda\lambda}^{(1)}(\vect{r}_A,\vect{r}_A,\mi\xi)\bigr]
\nonumber\\
=&\frac{\hbar}{2\pi\varepsilon_0}\!\int_0^\infty\!\!\!\dif\xi\,
 \alpha^\lambda(\mi\xi)
 \trace\ten{G}_{\lambda\lambda}^{(1)}(\vect{r}_A,\vect{r}_A,\mi\xi)
\end{align}
($\lambda=e,m$).

There are two important differences between the diamagnetic and the
paramagnetic magnetisabilities which will have an impact on the
associated potentials. Firstly, the diamagnetisability has an 
opposite sign with respect to the paramagnetisability, which is a 
consequence of the Lenz rule. Secondly, in contrast to the 
paramagnetisability which obeys the usual Kramers--Kronig relations,
the diamagnetisability is independent of frequency.
%
%
\subsection{Application: Atom in front of a perfectly reflecting
mirror}
\label{Sec3.2}
Let us consider an isotropic atom at distance $z_A$ from a perfectly
reflecting planar mirror. The magnetoelectric properties of the 
mirror are characterised by $\varepsilon\!=\!\infty$
($\mu\!=\!\infty$) for a perfectly conducting (infinitely permeable)
plate. The Green tensor reads \cite{0215}
\begin{multline}
\label{g1}
\bm{G}^{(1)}(\vect{r},\vect{r}',\mi\xi)\\
= \pm\frac{1}{8\pi^2}\int
\frac{\dif^2 q}
{b}\,\me^{\mi(\vect{q}_+\cdot\vect{r}-\vect{q}_-\cdot\vect{r}' )}
(\vect{e}_p^+\vect{e}_p^- -\vect{e}_s\vect{e}_s)
\end{multline}
($\vect{q}_\pm = \vect{q}\pm\mi b\vect{e}_z$, 
$\vect{q}\bot\vect{e}_z$, 
$q= |\vect{q}|$, $b=\sqrt{q^2+\xi^2/c^2}$) with the upper (lower) sign
corresponding to a perfectly conducting (infinitely permeable) plate
and the polarisation vectors $\vect{e}_s$ and $\vect{e}_p$ being
defined by ($\vect{e}_q = \vect{q}/q$)
\begin{equation}
\vect{e_s}=\vect{e}_q\vprod\vect{e}_z,\quad \vect{e}_p^\pm
=\frac{c}{\xi}(-\mi q\vect{e}_z \mp b\vect{e}_q).
\end{equation}
Evaluating the double curl of the Green tensor, we find that
$\ten{G}_{mm}^{(1)}(\vect{r},\vect{r}',\mi\xi)$ is equal to
$\!-\xi^2c^{-2}\ten{G}^{(1)}(\vect{r},\vect{r}',\mi\xi)$.
Substituting this into Eq.~(\ref{eq22}) and carrying out the
$\vect{q}$-integral, one finds
\begin{multline}
\label{up1}
U_d(z_A)=\frac{\pm\hbar\mu_0\beta^d}{16\pi^2z_A^3}\int_0^\infty
 \!\dif\xi\,\me^{-2z_A \xi/c}\\
 \times\left(1+2\,\frac{z_A\xi}{c}+2\,\frac{z_A^2\xi^2}{c^2}\right).
\end{multline}
After performing the $\xi$-integral, we find an
attractive (repulsive) CP potential
\begin{multline}
U_d(z_A) = \pm\frac{3\hbar\mu_0c\beta^d}{32\pi z_A^4} 
=\mp\frac{3\hbar\mu_0c}{32\pi z_A^4}\sum_{\alpha\in 
A}\frac{q_\alpha^2}
{6m_\alpha}\langle\hat{\bar{\vect{r}}}_\alpha^2\rangle .
\end{multline}
of a diamagnetic atom in front of a perfectly conducting (permeable)
plate. It is given by a single $1/z_A^4$ power law.

\begin{table*}[t]
\begin{center}
\begin{tabular}{|c|c||c|c|c|c|}
\hline
 \multicolumn{2}{|c||}{Plate $\rightarrow$}
 &\multicolumn{2}{c|}{Perfectly conducting}
 &\multicolumn{2}{c|}{Infinitely permeable}\\
\hline
\hspace{2ex}\,Atom $\downarrow$\hspace{2ex}\,&\hspace{2ex}\,Limit
$\rightarrow$\hspace{2ex}\,
 &\hspace{2ex}\,Retarded\hspace{2ex}\,
 &\hspace{2ex}\,Nonretarded\hspace{2ex}\,
 &\hspace{2ex}\,Retarded\hspace{2ex}\,
 &\hspace{2ex}\,Nonretarded\hspace{2ex}\,
\\ \hline\hline
\multicolumn{2}{|c||}{Electric}
 &\parbox{6ex}{$$-\frac{1}{z_A^4}$$}
 &\parbox{6ex}{$$-\frac{1}{z_A^3}$$}
 &\parbox{6ex}{$$+\frac{1}{z_A^4}$$}
 &\parbox{6ex}{$$+\frac{1}{z_A^3}$$}
 \\ \hline
\multicolumn{2}{|c||}{Paramagnetic}
 &\parbox{6ex}{$$+\frac{1}{z_A^4}$$}
 &\parbox{6ex}{$$+\frac{1}{z_A^3}$$}
 &\parbox{6ex}{$$-\frac{1}{z_A^4}$$}
 &\parbox{6ex}{$$-\frac{1}{z_A^3}$$}
 \\ \hline
\multicolumn{2}{|c||}{Diamagnetic}
 &\multicolumn{2}{c|}{\parbox{6ex}{$$-\frac{1}{z_A^4}$$}}
 &\multicolumn{2}{c|}{\parbox{6ex}{$$+\frac{1}{z_A^4}$$}}
 \\ \hline
\end{tabular}
\end{center}
\caption{
\label{Tab1}
Signs and asymptotic power laws of the ground-state CP potential of an
electric, para- or diamagnetic atom with a perfectly reflecting plate.
}
\end{table*}
In Table~\ref{Tab1}, we compare this result with the known findings
for electric and paramagnetic atoms. We recall that electric and
paramagnetic atoms interact with conducting and permeable plates
according to an 'equals-attract, opposites-repel' rule: An electric
plate attracts electric atoms while repelling (para)magnetic atoms,
with corresponding results for a (para)magnetic plate. In contrast to
this, diamagnetic potentials carry a sign that is opposite to their
paramagnetic counterparts. This is due to the Lenz rule as encoded in
the minus sign in the diamagnetic magnetisability~(\ref{eq21}). The
diamagnetic CP potential thus has the same sign as the corresponding
electric potential.

Another difference is the fact that the wavelengths of electric and
paramagnetic dipole transitions divide the CP potential into two
asymptotic regimes: the nonretarded regime of distances smaller than
these wavelengths and the opposite, retarded regime. The CP potential
follows two distinct $1/z_A^3$ and $1/z_A^4$ power laws in these
regimes, respectively. On the other hand, the frequency-independent
diamagnetic magnetisability lacks an intrinsic length scale. As a
result, the CP interaction with a perfectly reflecting plate follows a
retarded $1/z_A^4$ power law at all distances.
 
%
\section{Two-atom van der Waals interaction}
\label{Sec4}
Similar to the single-atom case, the body-assisted vdW force between
two atoms can be derived from the two-atom vdW potential
$U(\vect{r}_A,\vect{r}_B)$, which is that part of the energy shift
depending on the positions of both atoms,
\begin{equation}
\label{eq30}
U(\vect{r}_A,\vect{r}_B)=\Delta E(\vect{r}_A,\vect{r}_B).
\end{equation}
%
%
\subsection{Perturbation theory}
\label{Sec4.1}
Again, we assume the atom--field system to be in its uncoupled
ground-state $|0_A\rangle|0_B\rangle|\{0\}\rangle$, but now we have
to calculate the leading-order two-atom energy shift. We begin with
two purely diamagnetic atoms, in which case the vdW potential follows
from the second-order energy shift
\begin{equation}
\label{eq31}
\Delta_2 E=\sum_{\phi\neq 0}
 \frac{\langle 0|\hat{H}_{AF}^d+\hat{H}_{BF}^d|\phi\rangle
 \langle\phi|\hat{H}_{AF}^d+\hat{H}_{BF}^d|0\rangle}{E_0-E_\phi}\,.
\end{equation}
Note that the first order energy shift just leads to the sum of two
diamagnetic CP potentials as discussed in the previous
Sect.~\ref{Sec3.1}. Due the diamagnetic interaction Hamiltonian being
quadratic in the fields, the vdW shift already appears in second order
perturbation theory rather than in fourth order as for electric and
paramagnetic dipole transitions.

\begin{figure}[!t!]
\begin{center}
\includegraphics[width=1.0\linewidth]{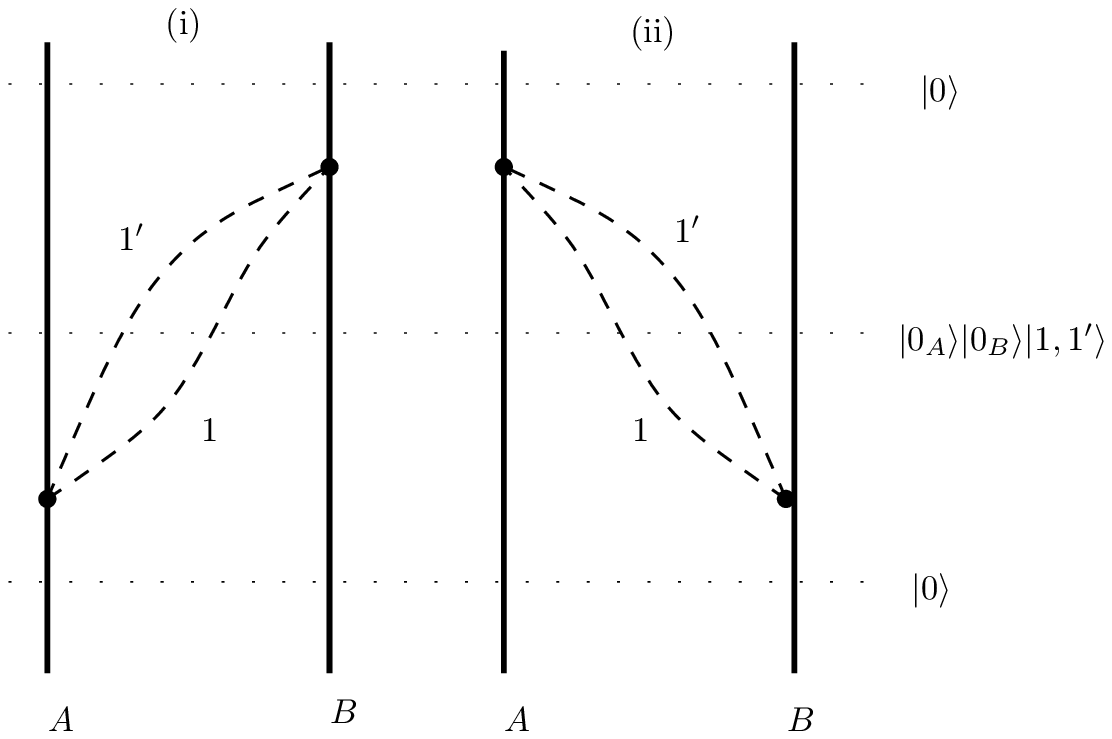}
\end{center}
\caption{
\label{Fig1}
Contributions to the vdW interaction of two diamagnetic atoms.
}
\end{figure}

The numerator of the term in Eq.~(\ref{eq31}), when read from
right to left, represents processes in which the system starts from
its ground state, goes to a state $|\phi\rangle$ due to a first
atom--field interaction, and finally returns to its ground state in
the course of a second interaction. As the interaction Hamiltonian is
quadratic in the magnetic field, the intermediate state $|\phi\rangle$
must involve the field in its ground state or exhibit two field
excitations (photons). Only the latter case leads to a genuine
two-atom interaction,
\begin{equation}
\label{phi}
|\phi\rangle = |0_A\rangle |0_B\rangle |1_{\lambda i}(\vect{r},
\omega),1_{\lambda'i'}(\vect{r}',\omega')\rangle
\end{equation}
with the two-photon state being defined by
\begin{gather}
\label{eq33}
|1_{\lambda i}(\vect{r},\omega),1_{\lambda'i'}(\vect{r}',
\omega')\rangle
=\tfrac{1}{\sqrt{2}}\hat{f}_{\lambda'i'}^\dagger(\vect{r}',
\omega')
\hat{f}_{\lambda i}^\dagger(\vect{r},\omega)|\{0\}\rangle.
\end{gather}
The photons must have been emitted by one of the atoms and then
absorbed by the other. This is schematically illustrated in
Fig.~\ref{Fig1}, where the solid lines and the dashed lines
represent the atoms and the photons, respectively (where time
progresses in the upwards direction). Note that the formal sum in
Eq.~(\ref{eq31}) involves sums over $\lambda$, $\lambda'$, $i$, $i'$
as well as integrals over $\vect{r}$, $\vect{r}'$, $\omega$ and
$\omega'$. We begin with the contribution $\Delta_2E_{(i)}$ to the
energy shift corresponding to Fig.~\ref{Fig1}(i), where atom $A$ emits
two photons and atom $B$ absorbs them, i.e., where the second matrix
element in the perturbative energy shift~(\ref{eq31}) is due to the
diamagnetic interaction of atom $A$ and the first one is due to that
of atom $B$. The required two-photon emission matrix element for atom
$A$ takes the form
%
\begin{multline}
\label{me1}
\langle\phi|\hat{H}_{AF}^d|0\rangle =\frac{-\beta_{Ajk}^d}
{2\sqrt{2}}\\
\times\langle\{0\}|\hat{f}_{\lambda'i'}(\vect{r}',
\omega')
\hat{f}_{\lambda i}(\vect{r},
\omega)\hat{B}_j(\vect{r}_A)\hat{B}_k(\vect{r}_A)|\{0\}\rangle,
\end{multline}
where Eqs.~(\ref{mag}), (\ref{phi}), and (\ref{eq33}) have been used. 
The two-photon absorption matrix element
$\langle 0|\hat{H}_{BF}^d|\phi\rangle$ can be obtained by taking the
complex conjugate of the above and replacing the labels $A$ by $B$.
After substituting Eq.~(\ref{eq7}) for the magnetic field and making
use of the commutation relations (\ref{comm1}) and (\ref{comm2}), the
respective term in Eq.~(\ref{eq31}) reads
\begin{multline}
\frac{\langle 0 |\hat{H}_{BF}^d|\phi\rangle
\langle\phi|\hat{H}_{AF}^d|0\rangle}{E_0-E_\phi}  = \frac{-1}
{2\omega^2\omega'^2\hbar(\omega+\omega')}\\
\times\trace\Big\{
\bm{\beta}_A^d\sprod\big[ 
\vect{\nabla}_{\!A}\vprod\ten{G}_{\lambda}^\ast(\vect{r}_A,\vect{r}
,\omega)\sprod\ten{G}_\lambda^\trans(\vect{r}_B,\vect{r},\omega)
\vprod\vect{\nabla}_{\!B}\big]\\
\cdot \bm{\beta}_B^d
\sprod\big[\vect{\nabla}_{\!B}\vprod\ten{G}_{\lambda'}(\vect{r}_B,
\vect{r}',\omega')\sprod
\ten{G}_{\lambda'}^{\ast \trans}(\vect{r}_A,\vect{r}',\omega')
\vprod\vect{\nabla}_{\!A}\big]\Big\}.
\end{multline}
At this stage, performing the integrals over $\vect{r}$ and 
$\vect{r}'$ included in Eq.~(\ref{eq31}) by means 
of the integral relation (\ref{eq14}) results in
\begin{multline}
 \label{DEi}\Delta_2 E_{(i)} = -\frac{\hbar\mu_0^2}
 {2\pi^2}\int_0^\infty\dif\omega \int_0^\infty\dif\omega'
\frac{1}{\omega+\omega'}\\
\times\trace\big[
\bm{\beta}_A^d\sprod \operatorname{Im}\,
 \ten{G}_{mm}(\vect{r}_A,\vect{r}_B,\omega)
 \cdot \bm{\beta}_B^d
\sprod\operatorname{Im}\,\ten{G}_{mm}(\vect{r}_B,\vect{r}_A,\omega')
\big]
\end{multline}
[Recall Eq.~(\ref{L})]. It can easily be seen that the contribution 
(ii) as depicted in Fig.~\ref{Fig1}(ii) is exactly the same. We hence
find
\begin{align}
\label{eq36}
&U_{dd}(\vect{r}_A,\vect{r}_B)
 =-\frac{\hbar\mu_0^2}{\pi^2}\int_0^\infty\dif\omega 
 \int_0^\infty\dif\omega'
\frac{1}{\omega+\omega'}\nonumber\\
&\times\trace\big[
\bm{\beta}_A^d\sprod \operatorname{Im}\,
\ten{G}_{mm}(\vect{r}_A,\vect{r}_B,\omega)
\cdot \bm{\beta}_B^d
\sprod\operatorname{Im}\,\ten{G}_{mm}(\vect{r}_B,\vect{r}_A,\omega')
\big].\nonumber\\
\end{align}
As for the single-atom potential, the result can be simplified via
contour-integral techniques. We first write
\begin{equation}
 \int_0^\infty\frac{\dif\omega'}{\omega+\omega'}\,
 \operatorname{Im}\ten{G}_{mm}(\omega') = 
 \operatorname{Im}\int_0^\infty\frac{\dif\omega'}{\omega+\omega'}\,
 \ten{G}_{mm}(\omega')
\end{equation}
and use the fact that the Green tensor is analytic in the upper half 
of the complex frequency plane including the real axis. Hence, we may
replace the integral with an integral along the positive imaginary
frequency axis and use the Schwarz reflection principle~(\ref{eq12})
to obtain
\begin{equation}
 \label{1stint}
\int_0^\infty\frac{\dif\omega'}{\omega+\omega'}\,
 \operatorname{Im}\ten{G}_{mm}(\omega') = 
 \int_0^\infty\dif\xi\,\frac{\omega}{\omega^2+\xi^2}\,
 \ten{G}_{mm}(\mi \xi).
\end{equation}
Substituting this result into Eq.~(\ref{eq36}), we next evaluate the
integral over $\omega$,
\begin{equation}
 \int_0^\infty\dif\omega\,\frac{\omega\operatorname{Im}\ten{G}_{mm}
 (\omega)}{\omega^2+\xi^2}\,
 =\operatorname{Im}\int_0^\infty\dif\omega\, 
 \frac{\omega\ten{G}_{mm}(\omega)}
 {\omega^2+\xi^2}\,
\end{equation}
The integrand in the right hand side has a simple pole at $\omega$ 
$\!=$ $\!\mi\xi$ in the upper half of the complex frequency plane. 
Again using Cauchy's theorem, we transform this integral into a
principal value integral along the positive imaginary axis, an
integral along an infinitesimal half circle around the pole, and a
(vanishing) integral along an infinite quarter circle. The first
integral is real, while  the second integral results in
$(\mi\pi/2)\ten{G}_{mm}(\mi\xi)$, so that
\begin{multline}
 \label{2ndint}
\int_0^\infty\dif\omega\,
\frac{\omega\operatorname{Im}\ten{G}_{mm}(\omega)}
{\omega^2+\xi^2}\,\\
=\operatorname{Im}
\bigg[\mathcal{P}\int_0^\infty\!\dif v\,\frac{v\ten{G}_{mm}(\mi 
v)}{v^2-\xi^2}
+\mi\,\frac{\pi}{2}\,\ten{G}_{mm}(\mi\xi)\bigg]\\
= \frac{\pi}{2}\,\ten{G}_{mm}(\mi\xi)
\end{multline}
($\mathcal{P}$: principal value). After substituting this result
together with Eq.~(\ref{1stint}) into Eq.~(\ref{eq36}), we find the
CP potential of two diamagnetic atoms,
\begin{align}
\label{eq97}
&U_{dd}(\vect{r}_A,\vect{r}_B)=-\frac{\hbar\mu_0^2}{2\pi}
 \int_0^\infty\dif\xi\nonumber\\
 &\hspace{6ex}\times \trace\bigl[\bm{\beta}_{A}^d
 \sprod
 \ten{G}_{mm}(\vect{r}_A,\vect{r}_B,\mi\xi)\sprod\bm{\beta}_{B}^d
 \sprod
 \ten{G}_{mm}(\vect{r}_B,\vect{r}_A,\mi\xi)
 \bigr]\nonumber\\
&=-\frac{\hbar\mu_0^2}{2\pi}
 \int_0^\infty\dif\xi\,\beta_{A}^d\beta_{B}^d
 \nonumber\\
 &\hspace{8ex}\times \trace\bigl[
  \ten{G}_{mm}(\vect{r}_A,\vect{r}_B,\mi\xi)
\sprod \ten{G}_{mm}(\vect{r}_B,\vect{r}_A,\mi\xi)
\bigr]
\end{align}
where the second equality holds for isotropic atoms.

Let us consider next the interaction of a diamagnetic atom $A$ with 
an electric atom $B$. It can be seen easily that the leading two-atom
energy shift is of third order in this case,
\begin{equation}
\label{3rd}
\Delta_3 E = \sum_{\phi,\psi\neq 0}\frac{\langle 0|
\hat{H}_{\mathrm{int}}|\psi\rangle\langle\psi|\hat{H}_{\mathrm{int}}|
\phi\rangle\langle\phi|\hat{H}_{\mathrm{int}}|0\rangle}{(E_\psi-E_0)
(E_\phi-E_0)}\,
\end{equation}
with
\begin{equation}
\hat{H}_{\mathrm{int}}\!=\hat{H}_{AF}^d\!+\hat{H}_{BF}^e\!= 
-\tfrac{1}{2}\hat{\vect{B}}
(\vect{r}_A)\sprod\hat{\bm{\beta}}^d_A\sprod\hat{\vect{B}}
(\vect{r}_A) - \hat{\vect{\mu}}_B\sprod\hat{\vect{E}}
(\vect{r}_B).
\end{equation}
\begin{figure}[!t!]
\begin{center}
\includegraphics[width=1.0\linewidth]{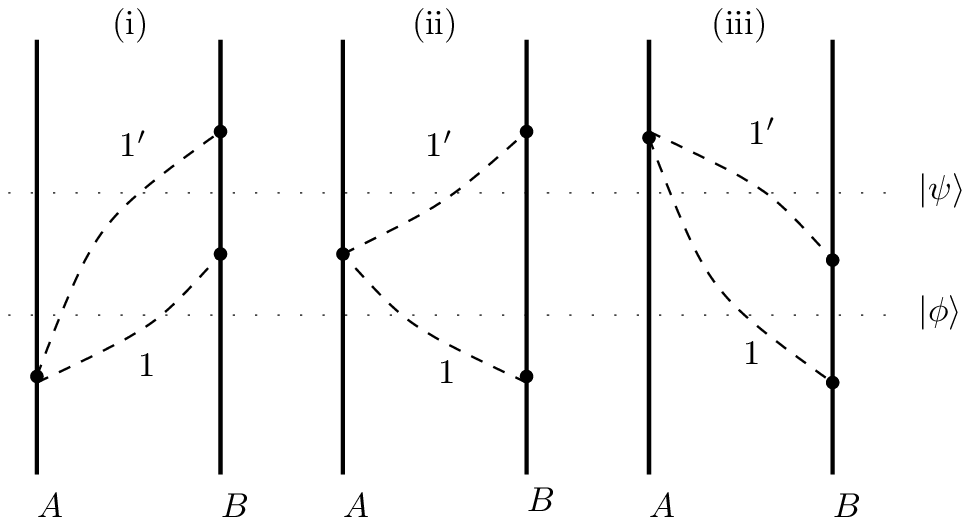}
\end{center}
\caption{
\label{Fig2}
Contributions to the vdW interaction of a diamagnetic atom $A$ with an
electric atom $B$.}
\end{figure}%
The relevant intermediate states $|\phi\rangle$ and $|\psi\rangle$
can easily be determined with the help of the diagrams in
Fig.~\ref{Fig2}. Let us begin with diagram~\ref{Fig2}(i) which
correspond to the the case where atom $A$ emits two photons and atom 
$B$ absorbs them one after another, so the relevant intermediate 
states are
\begin{eqnarray}
&&|\phi\rangle = |0_A\rangle|0_B\rangle|1_{\lambda i}(\vect{r},
\omega),1_{\lambda' i'}(\vect{r}',\omega')\rangle,\\
&&|\psi\rangle = |0_A\rangle|n_B\rangle|1_{\lambda'' i''}(\vect{r}'',
\omega'')\rangle.
\end{eqnarray}
In the two-photon emission matrix element, $\hat{H}_{\mathrm{int}}$ 
has to be replaced by $\hat{H}_{AF}^d$. This leads, as outlined below 
Eq.~(\ref{me1}), to
\begin{align}
\label{me2}
&\langle\phi|\hat{H}_{AF}^d|0\rangle\nonumber\\
&=
\frac{-1}{\sqrt{2}\omega\omega'}\ten{G}_\lambda^{\ast\trans}
(\vect{r}_A,\vect{r},
\omega)\vprod\overleftarrow{\vect{\nabla}}_{\!A}
\sprod{\bm{\beta}}_A^d\sprod
\vect{\nabla}_A\vprod\ten{G}_{\lambda'}^\ast(\vect{r}_A,\vect{r}',
\omega'),\nonumber\\
\end{align}
while in the two other matrix elements, $\hat{H}_{\mathrm{int}}$ 
has to be replaced with $\hat{H}_{BF}^e$. This yields
\begin{eqnarray}
\label{me3}
&&\hspace{-2ex}\langle\psi|-\hat{\vect{\mu}}_B\sprod\hat{\vect{E}}
(\vect{r}_B)
|\phi\rangle\nonumber\\
&&\hspace{-2ex}=\frac{-1}{\sqrt{2}}\Big\{\big[\vect{\mu}_{B}^{n0}
\sprod\ten{G}_{\lambda}(\vect{r}_B,\vect{r},
\omega)\big]_i\delta_{\lambda'\lambda''}\delta_{i'i''}
\delta(\vect{r}'\!
-\!\vect{r}'')\delta(\omega'\!-\!\omega'')\nonumber\\
&&\hspace{1ex}+\big[\vect{\mu}_{B}^{n0}\sprod
\ten{G}_{\lambda'}(\vect{r}_B,\vect{r}',
\omega')\big]_{i'}\delta_{\lambda\lambda''}\delta_{ii''}
\delta(\vect{r}
\!-\!\vect{r}'')\delta(\omega\!-\!\omega'')\Big\},\nonumber\\
\end{eqnarray}
\begin{equation}
\label{me4}
\langle0|-\hat{\vect{\mu}}_B\sprod\hat{\vect{E}}(\vect{r}_B)|
\psi\rangle = -\big[\vect{\mu}_{B}^{0n}\sprod
\ten{G}_{\lambda''}(\vect{r}_B,\vect{r}'',\omega'')\big]_{i''},
\end{equation}
where the expression (\ref{eq5}) for the electric field, as well as 
the commutation  relations (\ref{comm1}) and (\ref{comm2}) have been
used. Substitution of these matrix elements into Eq.~(\ref{3rd}) and
performing the integrals over $\vect{r}$, $\vect{r}'$, and
$\vect{r}''$ using the integral relation (\ref{eq14}), we find
the contribution of diagram (i) to the energy shift to be
\begin{multline}
\label{3rdE-i}
\Delta_3 E_{(\mathrm{i})} = \frac{\mu_0^2}
{\pi^2}\int_0^\infty\dif\omega\int_0^\infty\dif\omega' 
\frac{\omega\omega'}
{(\omega+\omega')(\omega_B^{n}+\omega)}\\
\times\big[\vect{d}_B^{0n}\sprod\operatorname{Im}
\ten{K}^\trans(\vect{r}_A,
\vect{r}_B,
\omega)\sprod\bm{\beta}_A^d
\sprod\operatorname{Im}\ten{K}(\vect{r}_A,\vect{r}_B,
\omega')\sprod\vect{d}_B^{n0}\big],
\end{multline}
where
\begin{equation}
\label{K}
\ten{K}(\vect{r},\vect{r}',
\omega)=\vect{\nabla}\vprod\ten{G}(\vect{r},\vect{r}',
\omega).
\end{equation}
The contribution to the energy shift from diagrams (ii) and (iii)
result in terms similar to Eq.~(\ref{3rdE-i}) except that the energy
denominator $(\omega$ $\!+\!$ $\omega')(\omega_B^{n}$ $\!+\!$
$\omega)$ has to be replaced with $(-\omega_B^{n}$ $\!-\!$
$\omega')(\omega_B^{n}$ $\!+\!$ $\omega)$ for diagram (ii)
and with $(\omega$ $\!+\!$ $\omega')(\omega_B^{n}$ $\!+\!$ 
$\omega')$ for diagram (iii). Summing all contributions, we find
\begin{multline}
\label{3rdE}
\Delta_3 E\
= \frac{2\mu_0^2}
{\pi^2}\int_0^\infty\!\dif\omega\!\int_0^\infty \frac{\dif\omega'\,
\omega_B^{n}\omega\omega'}
{(\omega_B^{n}+\omega)(\omega+\omega')(\omega_B^{n}+\omega')}\\
\times\big[\vect{d}_B^{0n}\sprod\operatorname{Im}
\ten{K}^\trans(\vect{r}_A,
\vect{r}_B,
\omega)\sprod\bm{\beta}_A^d
\sprod\operatorname{Im}\ten{K}(\vect{r}_A,\vect{r}_B,
\omega')\sprod\vect{d}_B^{n0}\big].
\end{multline}
By transforming the $\omega'$ integral by means of contour-integral 
to run along the positive imaginary axis,
\begin{multline}
\int_0^\infty \dif\omega'\frac{\omega'\operatorname{Im}\ten{K}
(\omega')}{(\omega+\omega')(\omega_B^{n}+\omega')}\\
=\int_0^\infty\dif\xi\frac{\xi^2(\omega_B^{n}+\omega)
\ten{K}(\mi\xi)}{(\omega^2+\xi^2)\big[(\omega_B^{n})^2+\xi^2\big]},
\end{multline}
and then performing the $\omega$ integral in the way explained above
Eq.~(\ref{2ndint}), one obtains
\begin{eqnarray}
\label{ude}
&&\hspace{-5ex}U_{de}(\vect{r}_A,\vect{r}_B)=\frac{\hbar\mu_0^2}
{2\pi}
 \int_0^\infty\dif\xi\,\xi^2\nonumber\\
&& \times
 \trace\bigl[\bm{\beta}_{A}^d
 \sprod\ten{K}(\vect{r}_A,\vect{r}_B,\mi\xi)
\sprod\bm{\alpha}_{B}(\mi\xi)
 \sprod
 \ten{K}^\trans(\vect{r}_A,\vect{r}_B,\mi\xi)\bigr]\nonumber\\
&&\hspace{-2ex}=\frac{\hbar\mu_0^2}{2\pi}
 \int_0^\infty\dif\xi\,\xi^2\beta_{A}^d\, \alpha_{B}
 (\mi\xi)\nonumber\\
&&\qquad\times \trace\bigl[\ten{K}(\vect{r}_A,\vect{r}_B,\mi\xi)
\sprod\ten{K}^\trans(\vect{r}_A,\vect{r}_B,\mi\xi)\bigr],
\end{eqnarray}
where the definition (\ref{eq25}) has been used and the second
equality holds for isotropic atoms. Our result (\ref{ude}) for the
vdW potential of a diamagnetic atom $A$ with an electric atom $B$
may be written in the form ($\bm{\beta}_A^d$ $\!\equiv$
$c^2\bm{\alpha}_A^d$)
\begin{eqnarray}
\label{ude2}
&&\hspace{-5ex}U_{de}(\vect{r}_A,\vect{r}_B)=
-\frac{\hbar}{2\pi\varepsilon_0^2}
 \int_0^\infty\dif\xi\,\nonumber\\
&& \times
 \trace\bigl[\bm{\alpha}_{A}^d
 \sprod\ten{G}_{me}(\vect{r}_A,\vect{r}_B,\mi\xi)
\sprod\bm{\alpha}_{B}^e(\mi\xi)
 \sprod
 \ten{G}_{em}(\vect{r}_B,\vect{r}_A,\mi\xi)\bigr]\nonumber\\
&&\hspace{-2ex}=-\frac{\hbar}{2\pi\varepsilon_0^2}
 \int_0^\infty\dif\xi\,\alpha_{A}^d\, \alpha_{B}^e
 (\mi\xi)\nonumber\\
&&\qquad\times \trace\bigl[\ten{G}_{me}(\vect{r}_A,\vect{r}_B,\mi\xi)
\sprod\ten{G}_{em}(\vect{r}_B,\vect{r}_A,\mi\xi)\bigr],
\end{eqnarray}
where 
\begin{align}
\label{Gem}
 &\ten{G}_{me}(\vect{r},\vect{r}',\omega) = \frac{\mi\omega}{c}
\ten{K}(\vect{r},\vect{r}',\omega)=\frac{\mi\omega}{c}
\vect{\nabla}\vprod\ten{G}(\vect{r},\vect{r}',\omega),\\
\label{Gme}
&\ten{G}_{em}(\vect{r},\vect{r}',\omega) = \frac{-\mi\omega}{c}
\ten{K}^\trans(\vect{r}',\vect{r},\omega)=\frac{\mi\omega}{c}
\ten{G}(\vect{r},\vect{r}',\omega)\vprod
\overleftarrow{\vect{\nabla}}'.
\end{align}

Obviously, the interaction between an electric atom $A$ with a
diamagnetic atom $B$ is given by the right hand side of
Eq.~(\ref{ude}) with the labels $A$ and $B$ being interchanged.

The vdW potential of a diamagnetic atom $A$ with a paramagnetic one
$B$ can be calculated in a completely analogous way. The lowest-order
energy-shift is again given by third-order perturbation theory with
the possible intermediate states given again as depicted in the
diagrams in Fig.~\ref{Fig2}. The result is ($\!\beta_A^p$ $\!\equiv$
$c^2\alpha_A^p$)
\begin{eqnarray}
\label{udp}
&&\hspace{-2ex}U_{dp}(\vect{r}_A,\vect{r}_B)=
-\frac{\hbar}{2\pi\varepsilon_0^2}
 \int_0^\infty\!\dif\xi\,\nonumber\\
&& \times
 \trace\bigl[\bm{\alpha}_{A}^d
 \sprod\ten{G}_{mm}(\vect{r}_A,\vect{r}_B,\mi\xi)
\sprod\bm{\alpha}^p_{B}(\mi\xi)
 \sprod
 \ten{G}_{mm}(\vect{r}_B,\vect{r}_A,\mi\xi)\bigr]\nonumber\\
&&\hspace{-2ex}=\frac{-\hbar}{2\pi\varepsilon_0^2}
 \int_0^\infty\!\dif\xi\,\alpha_{A}^d\, \alpha_{B}^p(\mi\xi) 
 \nonumber\\
&& \times\trace\bigl[\ten{G}_{mm}(\vect{r}_A,\vect{r}_B,\mi\xi)
\sprod\ten{G}_{mm}(\vect{r}_B,\vect{r}_A,\mi\xi)\bigr].
\end{eqnarray}

We have thus generalised the body-assisted vdW potential to allow for
atoms with a diamagnetic response. Recalling the previously derived
interaction of two paramagnetic atoms \cite{0831}, we again observe
that the generalization can be achieved by replacing the
paramagnetisability of each atom with the total magnetisability,
$\bm{\beta}^p(\omega)\mapsto\bm{\beta}(\omega)%
=\bm{\beta}^p(\omega)+\bm{\beta}^d$. The complete vdW potential of two
atoms with nontrivial electric, para- and diamagnetic properties can
thus be given as
\begin{equation}
\label{eq105}
U(\vect{r}_A,\vect{r}_B) =\sum_{\lambda,\lambda'=e,m}
U_{\lambda\lambda'}(\vect{r}_A,\vect{r}_B),
\end{equation}
with
\begin{eqnarray}
\label{ull}
&&\hspace{-2ex}U_{\lambda\lambda'}(\vect{r}_A,\vect{r}_B)=
-\frac{\hbar}{2\pi\varepsilon_0^2}
 \int_0^\infty\!\dif\xi\,\nonumber\\
&& \times
 \trace\bigl[\bm{\alpha}_{A}^\lambda(\mi\xi)
 \sprod\ten{G}_{\lambda\lambda'}(\vect{r}_A,\vect{r}_B,\mi\xi)
\sprod\bm{\alpha}^{\lambda'}_{B}(\mi\xi)
 \sprod
 \ten{G}_{\lambda'\lambda}(\vect{r}_B,\vect{r}_A,\mi\xi)\bigr]
\nonumber\\
&&\hspace{-2ex}=-\frac{\hbar}{2\pi\varepsilon_0^2}
 \int_0^\infty\!\dif\xi\,\alpha_{A}^{\lambda}(\mi\xi)\, 
\alpha_{B}^{\lambda'}(\mi\xi) 
 \nonumber\\
&&
\times\trace\bigl[\ten{G}_{\lambda\lambda'}(\vect{r}_A,\vect{r}_B,
\mi\xi)
\sprod\ten{G}_{\lambda'\lambda}(\vect{r}_B,\vect{r}_A,\mi\xi)\bigr],
\end{eqnarray}
Again, this implies that previous results for local-field corrected
vdW potentials \cite{0831} remain valid for diamagnetic atoms.
%
%
\subsection{Application: Two atoms in free space}
\label{Sec4.2}
As the simplest example for the two-atom interaction, let us consider 
two isotropic atoms $A$ and $B$ interacting with each other in free
space. The Green tensor reads (see e.g. Ref.~\cite{0853})
\begin{equation}
\label{G0}
 \ten{G}^{(0)}(\vect{r}_A,\vect{r}_B,\mi \xi) =
\frac{c^2}{4\pi \xi^2l^3}
\left[f(l\xi/c)\ten{I} - g(l\xi/c)\vect{e}_l\vect{e}_l\right]
\me^{-l\xi/c}
\end{equation}
($\vect{r}_A$ $\!\neq$ $\!\vect{r}_B$), where
$\vect{l}$ $\!=$ $\! \vect{r}_A$ $\!-$ $\!\vect{r}_B$,
$l=|\vect{l}|$, $\vect{e}_l$ $\!=$ $\!\vect{l}/l$, and
\begin{equation}
f(x) = 1+x+x^2,\quad g(x)=3+3x+x^2.
\end{equation}
Evaluating the curls as contained in definitions (\ref{L}),
(\ref{Gem}) and (\ref{Gme}), one easily finds
\begin{multline}
\ten{G}_{ee}^{(0)}(\vect{r}_A,\vect{r}_B,\mi\xi)
=\ten{G}_{mm}^{(0)}(\vect{r}_A,\vect{r}_B,\mi\xi)\\
=\frac{\xi^2}{c^2}\ten{G}^{(0)}
(\vect{r}_A,\vect{r}_B,\mi\xi),
\end{multline}
\begin{multline}
\ten{G}_{me}^{(0)}(\vect{r}_A,\vect{r}_B,\mi\xi)
=-\ten{G}_{em}^{(0)}(\vect{r}_A,\vect{r}_B,\mi\xi)\\
=\frac{\xi}{4\pi cl^2}(1+l\xi/c)\me^{-l\xi/c}\vect{e}_l\vprod\ten{I}.
\end{multline}
Substituting these into Eqs.~(\ref{ull}) and (\ref{eq105}) we obtain a
total free-space vdW potential
\begin{align}
\label{free-full}
&U(\vect{r}_A,\vect{r}_B) = U(l) =
\frac{\hbar\mu_0^2}{16\pi^3}\nonumber\\
&\times\bigg\{\frac{-1}{l^6}
\int_0^\infty\dif\xi\,[c^4\alpha_A(\mi \xi)\alpha_B(\mi \xi)
\!+\!\beta_A(\mi \xi)\beta_B(\mi \xi)]h_1(l\xi/c)\nonumber\\
&+\frac{1}{l^4}\int_0^\infty\dif\xi\,\xi^2
[\alpha_A(\mi \xi)\beta_B(\mi \xi)
\!+\!\beta_A(\mi \xi)\alpha_B(\mi \xi)]h_2(l\xi/c)\bigg\},\nonumber\\
\end{align}
where
\begin{equation}
h_1(x) =(3+6x+5x^2+2x^3+x^4)\me^{-2x},  
\end{equation}
\begin{equation}
h_2(x) =(1+2x+x^2)\me^{-2x}.
\end{equation}
As expected from the results of Sect.~\ref{Sec4.1}, this result has
the same form as the previously derived potential between electric
and paramagnetic atoms \cite{0831}, except that here the total
magnetisability appears in place of the paramagnetic one.

It is of particular interest to inspect the behaviour of the
interaction potential in the nonretarded/retarded limits, where the
atom--atom separation is small/large compared to the respective atomic
wavelengths. We focus here on the contribution of the atomic
diamagnetisabilities to the potential, given by Eq.~(\ref{free-full})
with $\beta(\mi\xi) \mapsto\beta^d$. 

The interaction potential between a diamagnetic atom $A$ and an
electric atom $B$, using the explicit expression for the
polarisability (\ref{eq25}) in Eq.~(\ref{free-full}), is found to be
an attractive potential in the form
\begin{equation}
\label{de-free}
U_{de}(l) = - \frac{\mu_0^2|\beta_A^d|}{24\pi^3l^4}
\sum_k\omega_B^{k}|\vect{\mu}_B^{0k}|^2\!\int_0^\infty\dif\xi\,
\frac{\xi^2}{\xi^2\!+\!{\omega_B^k}^2}\,h_2(l\xi/c).
\end{equation}
To achieve the limiting cases mentioned above, we note that in the
$\xi$-integral, $\omega_B^k$ $\!\ll$ $\!\xi(\omega_B^k$ $\!\gg$
$\!\xi)$ holds in the nonretarded (retarded) limit. This leads to a
$l^{-5}$-dependent and a  $l^{-7}$-dependent potential for the
nonretarded and retarded limits, respectively, as follows
\begin{align}
\label{de-nr}
&U_{de}^{\operatorname{n.r}}(l) =-
\frac{5\mu_0^2c|\beta_A^d|}{96\pi^3l^5}\sum_k\omega_B^{k}|
\vect{\mu}_B^{0k}|^2\,,\\
\label{de-r}
&U_{de}^{\operatorname{r}}(l) =- \frac{7\mu_0^2c^3|\beta_A^d|}{96
\pi^3l^7}\sum_k\frac{|\vect{\mu}_B^{0k}|^2}{\omega_B^{k}}\,\nonumber\\
&=-\frac{7\hbar\mu_0^2c^3}{64\pi^3l^7}\,|\beta_A^d|\alpha_B(0)\,.
\end{align}

The diamagnetic--paramagnetic part of the two-atom interaction in free
space is seen to be repulsive. It reads
\begin{equation}
\label{dp-free}
U_{dp}(l) =  \frac{\mu_0^2|\beta_A^d|}{24\pi^3l^6}
\sum_k\omega_B^{k}|\vect{m}_B^{0k}|^2\!\int_0^\infty\dif\xi\, 
\frac{1}{\xi^2\!+\!{\omega_B^k}^2}\,h_1(l\xi/c),
\end{equation}
where Eq.~(\ref{eq26}) is used for the paramagnetisability of 
atom $B$. For the nonretarded limit Eq.~(\ref{dp-free}) exhibits a
$l^{-6}$-dependence,
\begin{equation}
 U_{dp}^{\operatorname{n.r}}(l)
= \frac{\mu_0^2|\beta_A^d|}{16\pi^2l^6}\sum_k|\vect{m}_B^{k}|^2
= \frac{\mu_0^2|\beta_A^d|\langle\vect{m}_B^2\rangle}{16\pi^2l^6}\,,
\end{equation}
while in the retarded limit it tends to a $l^{-7}$-dependent 
potential,
\begin{equation}
U_{dp}^{\operatorname{r}}(l)=
\frac{23\mu_0^2c|\beta_A^d|}{96\pi^3 l^7}\sum_k
\frac{|\vect{m}_B^{0k}|^2}{\omega_B^k} = 
\frac{23\hbar\mu_0^2c}{64\pi^3 l^7}\,|\beta_A^d|\beta_B^p(0)\,.
\end{equation}

Finally, the diamagnetic--diamagnetic two-atom interaction in free
space shows a unique attractive $l^{-7}$-dependence for any arbitrary
range of atom--atom separation, due to the frequency-independence of
diamagnetisabilities. As can be obtained from Eq.~(\ref{free-full}),
it is given by
\begin{equation}
\label{dd-free}
U_{dd}(l)=-\frac{23\hbar\mu_0^2c}{64\pi^3l^7}\,\beta_A^d\beta_B^d\,.
\end{equation}

\begin{table*}[t]
\begin{center}
\begin{tabular}{|c|c||c|c|c|c|c|c|}
\hline
 \multicolumn{2}{|c||}{Atom $A$ $\rightarrow$}
 &\multicolumn{2}{c|}{Electric}
 &\multicolumn{2}{c|}{Paramagnetic}
 &\multicolumn{2}{c|}{Diamagnetic}\\
\hline
Atom $B$ $\downarrow$& Limit $\rightarrow$
 &\hspace{1ex}\,Retarded\hspace{1ex}\,
 &\hspace{1ex}\,Nonret.\hspace{1ex}\,
 &\hspace{1ex}\,Retarded\hspace{1ex}\,
 &\hspace{1ex}\,Nonret.\hspace{1ex}\,
 &\hspace{1ex}\,Retarded\hspace{1ex}\,
 &\hspace{1ex}\,Nonret.\hspace{1ex}\,
\\ \hline\hline
\multicolumn{2}{|c||}{Electric}
 &\parbox{7ex}{$$-\frac{1}{l^7}$$}
 &\parbox{7ex}{$$-\frac{1}{l^6}$$}
 &\parbox{7ex}{$$+\frac{1}{l^7}$$}
 &\parbox{7ex}{$$+\frac{1}{l^4}$$}
 &\parbox{7ex}{$$-\frac{1}{l^7}$$}
 &\parbox{7ex}{$$-\frac{1}{l^5}$$}
 \\ \hline
\multicolumn{2}{|c||}{Paramagnetic}
 &\parbox{7ex}{$$+\frac{1}{l^7}$$}
 &\parbox{7ex}{$$+\frac{1}{l^4}$$}
 &\parbox{7ex}{$$-\frac{1}{l^7}$$}
 &\parbox{7ex}{$$-\frac{1}{l^6}$$}
 &\parbox{7ex}{$$+\frac{1}{l^7}$$}
 &\parbox{7ex}{$$+\frac{1}{l^6}$$}
 \\ \hline
\multicolumn{2}{|c||}{Diamagnetic}
 &\parbox{7ex}{$$-\frac{1}{l^7}$$}
 &\parbox{7ex}{$$-\frac{1}{l^5}$$}
 &\parbox{7ex}{$$+\frac{1}{l^7}$$}
 &\parbox{7ex}{$$+\frac{1}{l^6}$$}
 &\multicolumn{2}{c|}{\parbox{7ex}{$$-\frac{1}{l^7}$$}}
 \\ \hline
\end{tabular}
\end{center}
\caption{
\label{Tab2}
Signs and asymptotic power laws of ground-state vdW potentials of
electric, para- or diamagnetic atoms in free space.
}
\end{table*}
In Table~\ref{Tab2}, we compare the signs and asymptotic power law of
the vdW potentials involving diamagnetic atoms with the known results
for purely electric or paramagnetic atoms \cite{0831}. We observe
that the replacement of a paramagnetic atom with a diamagnetic one
leads to a sign change of the vdW interaction as predicted by the
Lenz rule. In addition, the frequency-independence of the diamagnetic
magnetisability leads to new asymptotic power laws, e.g., $1/l^5$ for
the nonretarded electric--diamagnetic interaction. This result is
in between the $1/l^6$ and $1/l^4$ asymptotes found for the
electric--electric and electric--paramagnetic cases. 
%
%
\subsection{Comparison with Microscopic QED}
\label{Sec4.3}
It is instructive to compare the results obtained in this section
for the vdW dispersion interaction between atoms in free space when
either one or both species is diamagnetic using body-assisted fields,
with that derived using microscopic QED. In this second approach it is
common to use the well-established molecular QED theory
\cite{Craig84,Salam10}. For two mutually interacting particles $A$ and
$B$ in vacuum coupled to the radiation field, the total Hamiltonian is
written in the form of Eq.~(\ref{eq4}) with $\hat{H}_\xi$ given by
Eq.~(\ref{eq1}). In the multipolar coupling scheme, the Hamiltonian
operator for the free electromagnetic field is expressed as
\begin{multline}
\label{eq90}
\hat{H}_F=\frac{1}{2} \int\dif^3r\,
\biggl[\frac{1}{\varepsilon_0}\,\hat{\vect{d}}^2(\vect{r}) 
+\frac{1}{\mu_0}\,\hat{\vect{b}}^2(\vect{r})\biggr]\\
=\sum_{\vect{k},\lambda}\Bigl[\hat{a}^{(\lambda)\dagger}(\vect{k})
 \hat{a}^{(\lambda)}(\vect{k})+\tfrac{1}{2}\Bigr]\hbar\omega
\end{multline}
where $\hat{\vect{d}}(\vect{r})$ and $\hat{\vect{b}}(\vect{r})$ are
second quantized microscopic electric displacement and magnetic field
operators, respectively. They are commonly written as a mode sum in
terms of vacuum boson annihilation and creation operators
$\hat{a}^{(\lambda)}(\vect{k})$ and
$\hat{a}^{(\lambda)\dagger}(\vect{k})$ for a photon of wave vector
$\vect{k}$, polarization index $\lambda$, and circular frequency
$\omega=ck$, as in the second equality of Eq.~(\ref{eq90}). Analogous
to Eq.~(\ref{eq3}) with Eqs.~(\ref{elec-int})--(\ref{dia-int}), the
atom--field coupling Hamiltonian for the dispersion interactions of
interest is
\begin{equation}
\label{eq91}
\hat{H}_{\xi F}
 =-\varepsilon_0^{-1}\hat{\vect{\mu}}_\xi\sprod
 \hat{\vect{d}}(\vect{r}_\xi)
 -\hat{\vect{m}}_\xi\sprod\hat{\vect{b}}(\vect{r}_\xi)
 +\sum\limits_{\alpha\in\xi}
\frac{q_\alpha^2}{8m_\alpha}\bigl[\hat{\bar{\vect{r}}}_\alpha\vprod
\hat{\vect{b}}(\vect{r}_\xi)\bigr]^2
\end{equation}
It is worth noting the explicit appearance in the last two expressions
of the electric displacement field operator. This is a direct
consequence of adopting the multipolar framework, where the field
momentum canonically conjugate to the vector potential is proportional
to $\hat{\vect{d}}(\vect{r})$ instead of to the electric field itself.
This is a common feature of the microscopic and macrocopic approaches.
In the latter, the quantised field $\hat{\vect{E}}(\mathbf{r})$ also
refers to the Power--Zienau transformed electric field operator that
has to be regarded as displacement field with respect to the atomic
polarisation \cite{0853}.

Dispersion potentials between a diamagnetic atom and an
electrically polarizable and a magnetically susceptible atom, and
between two diamagnetic atoms may be computed in a manner similar to
that already detailed. As before, the last interaction is obtained via
second-order perturbation theory and is depicted in Fig.~\ref{Fig1},
while the former two occur in third-order and are described by
Fig.~\ref{Fig2}. Matrix elements are evaluated using unperturbed
product atom--field states $|n_\xi\rangle|m(\vect{k},\lambda)\rangle
=|n_\xi;m(\vect{k},\lambda)\rangle$  where a number state
representation is used to signify the number of photons present, in
this case $m$.

As expected, results identical to Eqs.~(\ref{de-free}),
(\ref{dp-free}) and (\ref{dd-free}) are found for the vdW dispersion
potentials between a diamagnetic and an electric atom, a diamagnetic
and a paramagnetic one, and between two diamagnetic atoms, valid for
the entire range of separation distance vector $\vect{l}$ beyond wave
function overlap of the two centres and extending out to infinity.
Explicit details of this calculation may be found in
Ref.~\cite{Salam00}, and the respective energy shifts are given by
Eqs.~(3.7), (3.20) and (3.43) of that paper. Furthermore, identical
near- and far-zone asymptotic limits clearly follow in each case. The
connection with the macroscopic approach can be established by
computing all necessary field expectation values and evaluating them
using the free-space Green tensor ~(\ref{G0}).
%
%
\section{Discussion and summary}
\label{Sec5}

We have calculated CP and vdW potentials of atoms with nontrivial
diamagnetic properties in the presence of arbitrary magnetoelectric
bodies on the basis of macroscopic QED and leading-order perturbation
theory. The nonlinear interaction generating the diamagnetic
interaction is quite different from the paramagnetic coupling and it
leads to different perturbative orders. Nevertheless, we have found
that diamagnetic atomic properties lead to dispersion potentials which
formally resemble those of paramagnetic atoms where the diamagnetic
magnetisability appears in place of the paramagnetic one. We have
explicitly shown this correspondence for one- and two-atom potentials,
but it is expected to hold for multi-atom vdW potentials as well.

However, the fact that the diamagnetic magnetisability is negative and
frequency-independent leads to diamagnetic potentials that differ in
signs and power laws from their paramagnetic counterparts.
Diamagnetic dispersion interactions carry the same sign as the
well-known electric potentials, which implies that diamagnetism alone
cannot be used to realise repulsive potentials. The unique power laws
resulting from the frequency-independence imply that diamagnetic
potentials have their strongest influence at short range.
%
%
\acknowledgments
We thank G. Barton for discussions. This work was supported by the
UK Engineering and Physical Sciences Research Council. 
%
%

\end{document}